\documentclass[12pt,preprint]{aastex}
\newcommand{\gapprox}   {\lower.4ex\hbox{$\;\buildrel >\over{\scriptstyle\sim}\;$}}
\newcommand{\lapprox}   {\lower.4ex\hbox{$\;\buildrel <\over{\scriptstyle\sim}\;$}}

\def\numden        {cm$^{-3}$}
\def\colden        {cm$^{-2}$}

\newcommand{\begeq}     {\begin{equation}}
\newcommand{\fineq}     {\end{equation}}

\newcommand{\Msun}      {\mbox{$\,M_{\mathord\odot}$}}

\newcommand{\Rsun}      {\mbox{$\,R_{\mathord\odot}$}}
\newcommand{\OMC}		{\mbox{${\rm O}-{\rm C}$}}

\newcommand{\Porb}{\mbox{$P_{\rm orb}$}}

\newcommand{\exo}{\mbox{EXO\,0748$-$676}}

\begin{document}

\title{Possible Magnetic Activity in the 
Low Mass X-ray Binary EXO~0748$-$676}

\author{Michael T. Wolff\altaffilmark{1},
Kent S. Wood\altaffilmark{2}}
\and
\author{ Paul S. Ray\altaffilmark{3}}

\altaffiltext{1}{Space Science Division, 
Naval Research Laboratory, Washington, DC 20375; Michael.Wolff@nrl.navy.mil}
\altaffiltext{2}{Space Science Division, 
Naval Research Laboratory, Washington, DC 20375; Kent.Wood@nrl.navy.mil}
\altaffiltext{3}{Space Science Division, 
Naval Research Laboratory, Washington, DC 20375; Paul.Ray@nrl.navy.mil}

\begin{abstract}

We report evidence of magnetic activity associated 
with the secondary star in the \exo\ low mass X-ray binary system. 
An analysis of a sequence of five consecutive X-ray 
eclipses observed during December 2003 with the {\it RXTE} satellite 
brings out a feature occurring during ingress we 
interpret as the X-ray photoelectric absorption shadow, 
as seen by an observer at Earth, 
of a plasma structure suspended 
above the surface of the secondary star.
The light curve feature consists of an initial drop 
in count rate to near zero (the absorption shadow) with a very 
short rebound to a significant fraction of the 
pre-ingress count rate and then a final plunge to totality over a
total time scale of $\sim$25 s.
The ingress feature persists 
for at least 5 consecutive orbital 
periods (a total of $\sim$19 hr), and possibly up to 5 days in our data.
Our data also show significant post-egress dipping during this 
eclipse sequence, unusual for this source, 
indicating possible secondary star mass ejection 
during this episode.

\end{abstract}


\section{Introduction}

\exo\ is the only persistently active low mass X-ray 
binary (LMXB) that displays full X-ray eclipses.
In order to observe orbital period variations in a 
LMXB we have conducted a long-term observational 
program \citep[see][and references therein]{whw+02} 
of monitoring the X-ray eclipses of \exo\ ($\Porb = 3.82$ hr) with 
the {\it Rossi X-Ray Timing Explorer} ({\it RXTE}) satellite. 
By monitoring the mid-eclipse times we can investigate changes 
in the orbital period and constrain the magnitudes of 
the physical processes that cause the observed changes, allowing
estimation of their long-term evolutionary effects.
 
In this letter we report on a chance observational circumstance 
that has allowed us to identify a possible X-ray signature of magnetic 
activity in the secondary star in \exo.
Magnetic activity has been observed for some time in both
the white dwarfs and secondary stars of cataclysmic variable 
systems \citep[e.g.,][]{warn95,ws01}.
To our knowledge, however, magnetic activity associated with 
the secondary stars of neutron star or black hole candidate 
LMXB systems has yet to be reported.
Indeed, the secondary star in the \exo\ system has never been reliably 
observed because of the dominance of the optical emission by the
accretion disk \citep{sc87,phs+06}.
Magnetic fields associated with the companion star may lead to 
modulation of rates of angular momentum exchange and mass transfer 
between the binary components and changes in the structure of 
the secondary star \citep{tv06}.
Estimates of the magnitude of magnetic effects on LMXB evolution 
have been hampered by the lack of observations 
of magnetic activity in real systems that can be 
tied directly to the observable properties of the secondary star.
This is because there are few systems where distinct evolutionary 
processes can be directly observed (e.g., observing orbital
evolution by timing eclipses) and even in systems where 
such observations are possible the nature of these processes 
is controversial \citep[e.g., see discussion of several eclipsing LMXBs in][]{whw+02}.
Finally, in analogy with the 11-year solar cycle, 
magnetic fields associated with the secondary star in a LMXB  
could show activity cycles \citep{lrr98}, and may also 
cause observable transient phenomena.

\section{{\it RXTE} Observations}

The observations we report here were all made with the
Proportional Counter Array (PCA) on the {\it RXTE}.
The PCA is an array of five large-area X-ray proportional counters
(Proportional Counter Units or PCUs) with microsecond timing accuracy 
\citep{jmr+06}.  
From the beginning of the {\it RXTE} mission (1996) we have 
monitored \exo\ to time X-ray eclipses with 
sub-second accuracy.
This monitoring is organized into ``campaigns'' consisting 
of 4$-$6 separate observations of \exo\ over roughly one day with 
one or two month spacing between campaigns.
Our timing accuracy for each eclipse is limited by 
counting statistics in the X-ray flux and any intrinsic variability in 
the X-ray emission.  

Eclipse cycle ($N$) is determined by the
numbering system of \citet{psvc91} with the updated ephemeris:
$T_0 ({\rm TDB;MJD}) = 46111.0751910$ and $\Porb = 0.15933778478$ days. 
We record full photon event data using {\it GoodXenon} mode.
For each eclipse light curve we select only layer 1 photon events from 
those PCUs that are on during the entire observation.
This allows us to minimize the noise in background-subtracted light 
curves at very low count rates (e.g., at eclipse totality where 
the source count rate is near zero). 
However, below we will have occasion to refer to the hardness of
the source spectrum and in those situations we select events from
all layers allowing us to bring out more clearly the spectral character 
at energies above 7 keV of the ingress feature we study.
Because of improvements in the PCA background model we extract events
in the energy range 2$-$20 keV, rather than 
2$-$12 keV as done in \citet{whw+02}
We utilize the combined background model released on
August 6, 2006 available at the {\it RXTE} web 
site\footnote{PCA X-ray background model information is 
found at the web site http://rxte.gsfc.nasa.gov/docs/xte/pca\_news.html.}.
Each {\it RXTE} observed eclipse is processed using 
the FTOOLS data analysis package. 
See \citet{whw+02} for further details of our analysis procedure.

\section{Observed Eclipse Profiles}

\begin{figure}
\centerline{\includegraphics[height=5.9in,angle=0.0]{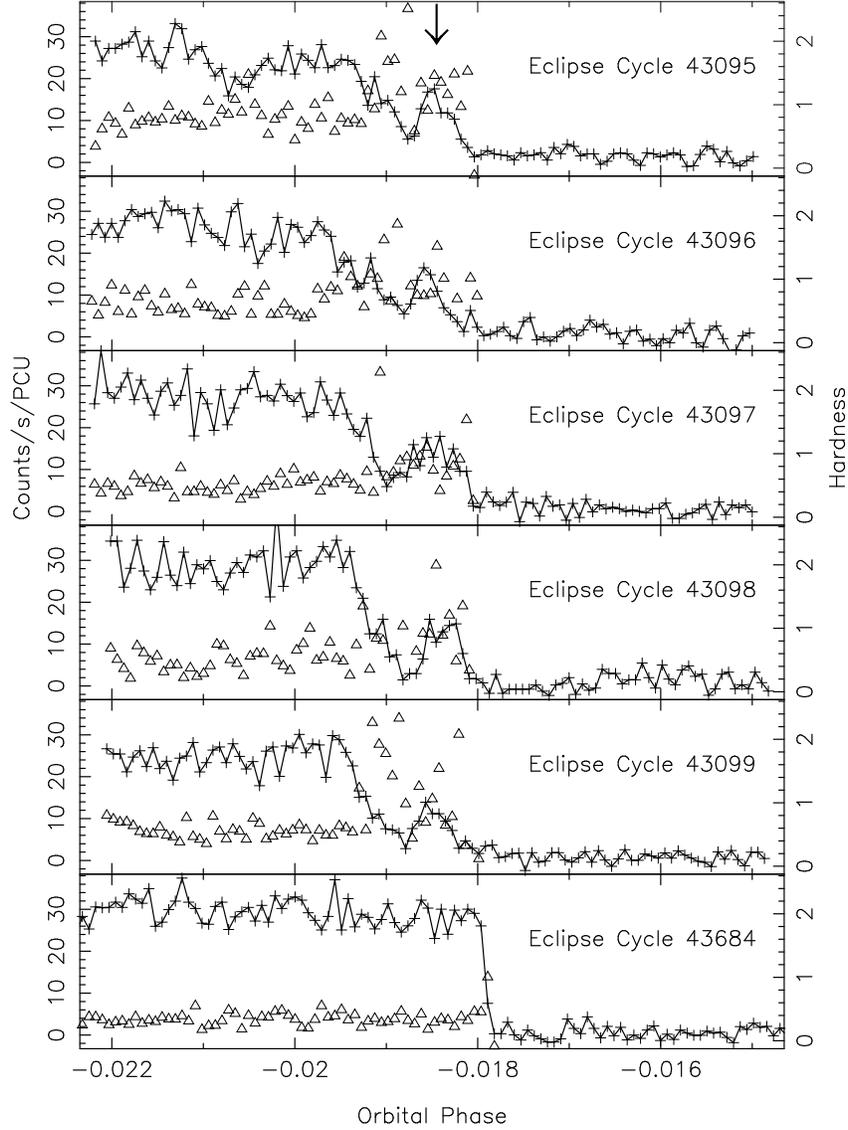}}
\caption{Six panels that show the light curves (crosses) and hardness
ratios (triangles) of the five spiked eclipse ingresses  
(top five panels), and one particularly 
sharp eclipse ($N = 43684$) observed 93 days later. 
Phase zero is set by the mid-eclipse 
time for eclipse $N = 43684$ and then applied to the other eclipses
so that all ingress points can be seen phased to the same
{\it local} ephemeris.
The light curve points are the PCA count rate per PCU in
0.5 s time bins for the 2$-$20 keV energy band and are connected 
with lines to allow them to be easily distinguished.
The hardness is the ratio of the count rate in the 7$-$20 keV energy
band to that in the 2$-$7 keV band.
Inside eclipse totality hardness points are not plotted.
The spike we discuss is centered at phase $\phi = -0.0185$ (downward arrow) 
while the real X-ray ingress appears to start 8 s later 
at phase $\phi = -0.0179$.
\label{fig:sixpanel}}
\end{figure}

Of primary interest to us here are five sequential 
eclipses, $N = 43095$ through $N = 43099$, observed 
by {\it RXTE} on December 4$-$5, 2003. 
These are listed in Table~\ref{tbl:humpeclipses} where we 
also include for comparison eclipses occurring both before and after 
this sequence.
The ingress profiles for these five eclipses are shown 
in Figure~\ref{fig:sixpanel} along with, for comparison, 
the ingress of an eclipse observed roughly 93 days later ($N = 43684$).
All five eclipse ingresses ($N = 43095 - 43099$) have an 
unusual structure, namely, as X-ray eclipse first contact is approached, 
the count rate first goes down, then rebounds to 
a significant fraction of the pre-ingress level (we refer to this 
as the ingress ``spike''), then drops to  
a count rate of essentially zero as real totality begins.
Contrast this behavior with the eclipse shown in the bottom panel of 
Figure~\ref{fig:sixpanel} ($N = 43684$).
In that case the ingress is sharp, lasting only about 2$-$3 seconds 
whereas in each of the $N = 43095 - 43099$ eclipses the complete 
ingress structure lasts at least 25 s.
This spike feature lasts for at least five orbits ($\sim$19 hr) as shown
in Figure~\ref{fig:sixpanel}.
Also, the ingress spikes occur at the same phase of the orbit relative 
to the occulting edge of the secondary for each 
$N = 43095 - 43099$ eclipse. 
The physical cause of the light curve spike 
occurs on the same side of the secondary for each of the five orbits.
Furthermore, examining eclipse $N = 43065$ we
see the ingress comes $\sim$10 s early compared to a local
constant period ephemeris, a possible sign of the emergence of the 
loop from the stellar interior before its geometrical 
structure allows X-rays to partially transit through it on 
their way to the Earth. 

The energy distribution of the photons in the 
ingress spike is ``hard'' in that
the ratio of hard ($E > 7$ keV) to soft ($E < 7$ keV) count rate 
goes up in the spike feature and the extended ingress, as can again 
be seen in Figure~\ref{fig:sixpanel}.
Indeed, above 7 keV the eclipse ingress, while somewhat disorganized, 
{\it occurs on the totality-side of the spike}.
In other words, the initial count rate drop occurs because
of soft photons being removed from the X-ray beam along our line of sight.
This is a signature of photoelectric absorption by material moving ahead of  
the occulting edge of the secondary star. 

\begin{figure}
\centerline{\includegraphics[height=6.0in,angle=0.0]{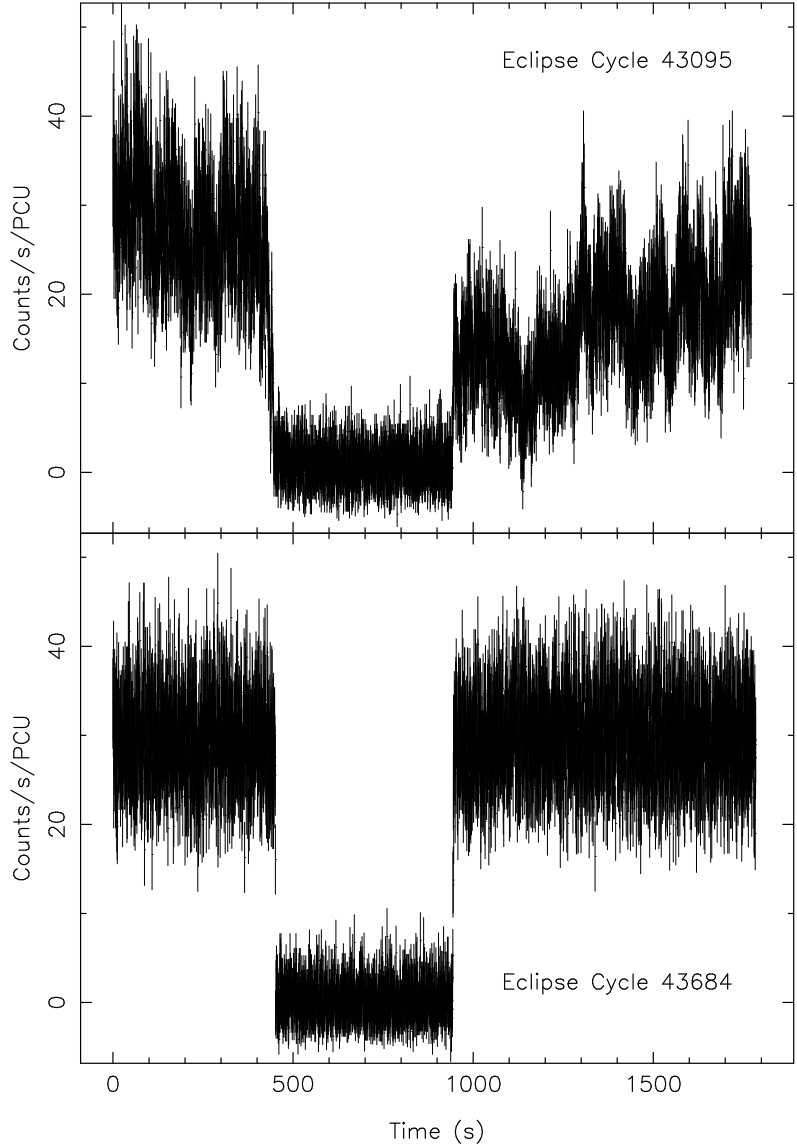}}
\caption{The light curves for the 2$-$20 keV energy band for the region 
around the eclipses $N = 43095$ (top panel), and $N = 43684$ (bottom panel).
Eclipse $N = 43095$ is the first of the eclipses with
ingress spikes and eclipse $N = 43684$ occurs three months later.
The binary orbit phasing is arbitrary, only layer 1 events are 
included with time binning 0.5 s.
Note the strong dipping after $N = 43095$ eclipse egress whereas
the post-egress count rate is steady for the $N = 43684$ eclipse. 
Our experience is that this level of post-egress dipping is rare 
during our monitoring observations.
The count rate in each time bin is plotted with $\pm 1 \sigma$ error bars
and the time scale is compressed relative to Figure~\ref{fig:sixpanel} 
hence the ingress spike feature shown in Figure~\ref{fig:sixpanel} does not 
stand out as strongly in the upper panel of this figure. 
\label{fig:lc70048-13-16-00}}
\end{figure}

Another feature of this set of eclipses is a persistent
level of {\it post-egress dipping} in the PCA energy band 
shown in Figure~\ref{fig:lc70048-13-16-00}.
Based on our experience from the eclipse monitoring program, 
even though the PCA energy band is relatively hard, 
dipping before ingress is not unusual.
In fact, dipping at all phases of the \exo\ orbit has been 
reported by several authors \citep[e.g.,][]{cbda98,spo05} 
but dipping of this magnitude at the immediate post-egress
orbit phase is relatively rare.
Examining the spectral fits to the post-egress PCA data shows,
assuming solar abundances, 
the inferred hydrogen column density is $n_H \sim 10^{23}$ \colden\ 
for these post-eclipse data.
\citet{tcs+97} found from {\it ASCA} observations of \exo\
that $n_H$ varied from $4 \times 10^{21}$ \colden\ outside of 
any dipping to a high of $10^{23}$ \colden\ during deep dips.
After the five eclipse sequence of interest the observed column 
density in our post-egress spectral fits goes down by more than 
an order of magnitude and the strong post-egress dipping ceases.

\section{Discussion}

\begin{figure}
\centerline{\includegraphics[height=3.0in,angle=0.0]{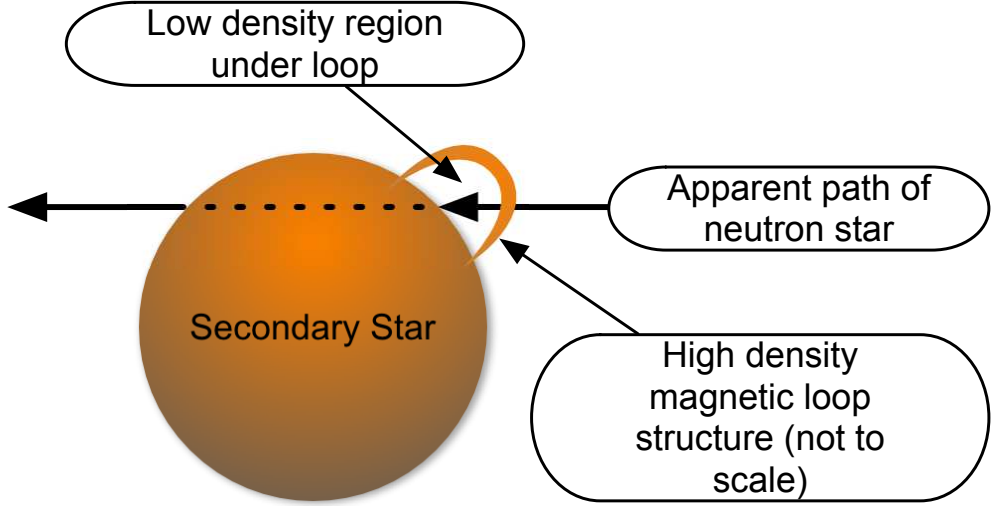}}
\caption{A schematic drawing of our model for the ``spike'' eclipses
from the point of view of an observer at Earth. 
The spike in the ingress profile corresponds to when the apparent 
neutron star position is between the loop and the edge of the secondary star and 
shinning through the lower density material under the loop.
True eclipse totality begins as the apparent position of the neutron star moves 
completely behind the secondary star and lasts until neutron star re-emergence. 
\label{fig:cartonpicture}}
\end{figure}

The spiked ingress behavior described above can occur if, as the 
line of sight from the neutron star to the Earth comes close to the 
occulting edge of the secondary star, a high-density 
plasma structure suspended above the surface of the secondary and in 
the plane of the line of sight, absorbs X-ray photons out of 
the beam. 
Then, a few seconds later the X-ray line of sight passes 
through a lower density region between the suspended structure 
and the true occulting edge of the secondary star, 
creating the brief spike in the light curve.
This eclipse geometry is schematically shown in Figure~\ref{fig:cartonpicture}.
Such a structure might be similar to a magnetic loop in the solar
atmosphere that is anchored at two foot points of opposite magnetic 
polarity \citep[e.g.,][]{bll+02}.

Structures of the sort we propose may last for several rotations 
of the secondary star because the secondary's magnetic field along
with the relatively gentle gravitation potential gradient 
will maintain the mechanical rigidity of the structure.
Utilizing Kepler's law and the \exo\ orbital period of \Porb = 3.82 hr, 
we obtain
$ A \,\, = \,\, 1.4 \left( \frac{M_1}{1.4 \Msun} \right) ( 1 + q )^{1/3} \, \Rsun, $
for the binary separation, where $q = M_2/M_1$ is the binary 
mass ratio and $M_1$ is the compact object mass.
We assume component (1) is a neutron star and set $M_1 = 1.4 \Msun$.
If $M_2 = 0.4 \Msun$ \citep{pwgg86} then $q = 0.286$ 
and $A = 1.5$ \Rsun.
The radius of the secondary star, $R_2$, is
the radius of the Roche lobe surrounding it \citep{egg83},  
\begin{equation}
R_2 \,\, = \,\, \frac{0.49 q^{2/3}}{0.6 q^{2/3} + 
\ln ( 1 + q^{1/3})} \, A \, \sim \, 0.42 \, \Rsun,
\label{eq:roche}
\end{equation}
for the above $A$ and $q$.
An estimate of the width of the feature is
$w \sim \frac{10 s}{\frac{1}{2} \Delta T_{ec}} \times R_2 \sim 0.0084 R_2 \sim 5.9 \times 10^8 \, \mathrm{cm}$
where $\Delta T_{ec}$ is the average duration of eclipse 
totality (495 s), 10 s is the time scale of the drop 
before the spike feature,
and where we have assumed that the system 
inclination is $i \geq 75^{\circ}$ \citep{pwgg86}. 
From Figure~\ref{fig:sixpanel} the count rate goes down 75\% 
during the initial drop before the spike. 
Simulating the \exo\ pre-ingress spectrum to determine the necessary 
column density to make the 2$-$7 keV flux drop by this amount 
yields a column density $N_H \sim 2.5 \times 10^{23}$ \colden\ implying
a hydrogen density $n_H \sim N_H / w \sim 4 \times 10^{14}$ \numden\ for 
the loop structure.
This density is higher than the corresponding densities in
solar magnetic loops \citep[$n_H \sim 10^9$\ \numden;][]{bll+02}
and even higher than coronal densities observed in the 
G5III flare system Capella 
\citep[$n_H \sim 4 \times 10^{11} - 10^{13}$\ \numden;][]{dbd+93}.
The secondary mass we use above ($M_2 = 0.4 \Msun$) corresponds to
either a late K or early M type main sequence star. 
Such stars can have magnetic activity if they rotate and
if they have significant convective envelopes \citep{hn87}. 
The fact that \exo\ secondary rotates very fast (assuming it 
rotates synchronously) and that stars in this mass range are 
thought to have significant convective layers does support 
our magnetic loop hypothesis.

Figure~\ref{fig:lc70048-13-16-00} shows that post-egress dipping is 
prominent during the five eclipse sequence we are studying.
This post-egress dipping could indicate significant mass ejection 
that moves away from the secondary accompanying the magnetic activity.
Such mass ejections could be similar to coronal mass ejections that 
sometimes accompany solar magnetic activity.
Because we observe only a total column density, however,
we can not establish whether this material is associated 
with the accretion disk inside the neutron star Roche lobe,
or it is outside the neutron star Roche lobe and moving 
away radially from the secondary.

In the simple model of \citet{ll02}, in order for the structures to
last in the solar corona they must be in hydrostatic equilibrium, 
internal velocities within the loop must be negligible, pressure 
gradients should be small, and energy dissipation processes 
within the loop must be uniform.
Because the loop structure resides near the surface of the 
secondary star it will be near a saddle point in the Roche 
gravitational potential function.
Thus, the effective force on a plasma structure will be 
small near the surface of the secondary. 
Second, if the forces on the loop are small then, unless there is 
significant non-uniformity in the energy dissipation or heating 
processes along the loop, 
velocities should be negligibly 
small and any pressure gradients should also be small.
However, heating and cooling processes within the loop are 
difficult to characterize.
The loop may have a hydrogen-poor abundance reflecting 
the abundance of the secondary star.
Also, the loop will be bathed in 
substantial X-rays from the accretion onto the neutron star, 
the radiation from an optically thin coronal plasma 
(if one exists), and from the radiation emerging from the 
stellar photosphere.
At a density of $n_H \sim 4 \times 10^{14}$ \numden\ the loop may  
be optically thick, making its energy balance particularly difficult 
to determine. 

The set of anomalous eclipses runs from cycle 
$N = 43095$ to $N = 43099$.
Based on the \OMC\ residuals there is considerable intrinsic 
period jitter of the type studied in \citet{hwc97} 
and \citet{whw+02} prior to the spike eclipses.
Other than this jitter, however, both before and after this set of 
observations the eclipses appear to be normal in that both the ingresses 
and egresses are well behaved although some eclipses 
have longer than average duration.
However, the eclipses running from $N = 43684$ to $N = 44363$,
in particular the set $N = 43684- 43687$ 
(Table~\ref{tbl:humpeclipses}), after the ingress feature disappears, 
are among the most stable eclipses we have observed from \exo\ 
with {\it RXTE} over our entire monitoring program.
The emergence of the ingress magnetic loop and significant 
post-eclipse dipping may signify a dramatic transient event followed 
by quiescence in the system.
The stable eclipses $N = 43684 - 43687$ occur during the first of 
a series of luminosity excursions observed by the {\it RXTE} 
All Sky Monitor that have continued on and off to the present day.
During those excursions the \exo\ X-ray brightness can increase from its
average of $\sim 8$ mcrab by factors of $2-5$ for weeks at a time.
Magnetic activity in the secondary star may be modulating 
the mass accretion rate onto the neutron star and affecting
the system X-ray luminosity.

\section{Conclusions}

We have reported on a sequence of five consecutive X-ray 
eclipses observed by {\it RXTE} during December 2003 from 
the \exo\ LMXB system that show a repeating feature during 
ingress transitions.
This feature is an anomalous X-ray peak during ingress we 
interpret as X-rays emitted from near the neutron star passing 
{\it underneath} a rigid plasma structure suspended above 
the surface of the secondary. 
This loop structure may be similar to magnetic loops 
suspended above the solar photosphere.
This conclusion is based on the ingress feature lasting for at least five 
orbit periods and perhaps as long as five days, although this latter 
conclusion depends on the interpretation of one ingress profile.
The occurrence of this ingress feature may be an indication of a 
transient event of significant magnitude in the secondary star 
since the \OMC\ residuals around the time of the event
show the mid-eclipse timings wandering before the ingress feature 
appears but settling down after the ingress feature disappears.
During this sequence of eclipses we also observed significant 
{\it post}-egress dipping which, as we noted, is unusual.
This suggests that significant mass may have been liberated from 
the secondary star (similar to a coronal mass ejection) as part
of this transient event.  
The occurrence of the X-ray spike on the limb of the secondary at
the same orbit phase in sequential eclipses 
is consistent with the conventional picture of synchronous rotation 
as suggested by a number of investigations \citep[e.g.,][]{vp95}.
Magnetically maintained features playing a role in 
determining the eclipse occulting edge geometry 
could lead to significant departures from symmetry between 
ingress and egress.
This may be the cause of the intrinsic jitter identified by
\citet{hwc97}\ and \citet{whw+02}.
If true then the magnetic field of the secondary may
perturb the structure of the secondary star, moderating both mass transfer 
and angular momentum exchange.

\begin{acknowledgements}
We thank Drs. Paul Hertz, Gerald Share, and J. Martin Laming 
for useful discussions and Evan Smith for help 
with {\it RXTE} scheduling.
We thank an anonymous referee for critically reading the manuscript
and many helpful suggestions.  
This work was supported by the NASA {\it RXTE} Guest Observer Program 
and the Office of Naval Research.
\end{acknowledgements}


%
%
\begin{deluxetable}{cccc}
\tablecolumns{4}
\tabletypesize{\scriptsize}
\tablecaption{\bf{{\it RXTE} Timing of Selected Full \exo\ X-ray Eclipses}
\label{tbl:humpeclipses}}
\tablewidth{420pt}
\tablehead{\colhead{{\it RXTE} ObsID} &\colhead{Eclipse Cycle} 
&\colhead{Fitted Mid-Eclipse Time\tablenotemark{1}}
 &\colhead{Comment}\\  &  & (MJD;TDB) & }
\startdata
70048-13-12-00   &42716  &52917.347911(3)  &Normal eclipse.\\
70048-13-13-00   &42717  &52917.507251(5)  &Burst partially obscures post-egress light curve.\\
70048-13-14-00   &42718  &52917.666590(6)  &Normal eclipse.\\
70048-13-15-00   &42719  &52917.825925(5)  &Normal eclipse; Some post-egress dipping.\\
80040-01-07-00   &43065  &52972.956764(5)  &Apparent eclipse duration 505 s.\\
70048-13-16-00   &43095  &52977.736972(7)\tablenotemark{2} &Double-peaked ingress; Post-egress dipping.\\
70048-13-17-00   &43096  &52977.89628(1)\tablenotemark{2}  &Double-peaked ingress; Post-egress dipping.\\
70048-13-18-00   &43097  &52978.055689(4)\tablenotemark{2} &Double-peaked ingress; Post-egress dipping.\\
70048-13-19-00G  &43098  &52978.214950(7)\tablenotemark{2} &Double-peaked ingress; Post-egress dipping.\\
70048-13-20-00   &43099  &52978.374306(2)\tablenotemark{2} &Double-peaked ingress.\\
80040-01-06-00   &43342  &53017.093350(3)  &Normal eclipse.\\
90059-02-01-00   &43684  &53071.5869083(9)  &Eclipse features sharp; High count rate.\\
90059-02-02-00   &43685  &53071.746244(3)  &Eclipse features sharp; High count rate.\\
90059-02-03-00   &43686  &53071.9055842(6)  &Eclipse features sharp; High count rate.\\
90059-02-04-00   &43687  &53072.064922(2)  &Eclipse features sharp; High count rate.\\
\enddata
\tablenotetext{1}{The fitted mid-eclipse times given to the first 
uncertain digit; The errors are in parenthesis for that digit.}
\tablenotetext{2}{No adjustment is made in the fitted mid-eclipse 
time for the double-peaked ingress profile.}
\end{deluxetable}

\end{document}